\documentclass{aa}
\usepackage{graphicx}

\begin{document}

\newcommand{\lsim}{\mathrel{\rlap{\raise -.3ex\hbox{${\scriptstyle\sim}$}}%
                   \raise .6ex\hbox{${\scriptstyle <}$}}}%
\newcommand{\gsim}{\mathrel{\rlap{\raise -.3ex\hbox{${\scriptstyle\sim}$}}%
                   \raise .6ex\hbox{${\scriptstyle >}$}}}%

\title{Extragalactic source contributions to arcminute-scale Cosmic Microwave Background anisotropies}
\titlerunning{Arcminute-scale anisotropies}
\author{Luigi Toffolatti\inst{1} \and Mattia
Negrello\inst{2} \and Joaqu{\'\i}n Gonz\'alez-Nuevo\inst{1}  \and
Gianfranco De Zotti\inst{2,3} \and Laura Silva\inst{4} \and Gian
Luigi Granato\inst{2,3} \and Francisco Arg\"ueso\inst{5}}

   \offprints{L. Toffolatti: toffol@pinon.ccu.uniovi.es}

\institute{ Departamento de Física, Universidad de Oviedo, c.
Calvo Sotelo s/n, 33007 Oviedo, Spain\\
   \and International School for Advanced Studies,
SISSA/ISAS, Via Beirut 2-4, I-34014 Trieste, Italy\\
\and INAF--Osservatorio Astronomico di Padova, Vicolo
dell'Osservatorio 5, I-35122 Padova, Italy\\
\and INAF--Osservatorio Astronomico di Trieste, Via Tiepolo 11,
I-34131 Trieste, Italy\\
\and Departamento de Matem\'aticas, Universidad de Oviedo, c.
Calvo Sotelo s/n, 33007 Oviedo, Spain
    }

\date{Received / Accepted }







\abstract{The possible contributions of the various classes of
extragalactic sources (including, in addition to the canonical
radio sources, GHz Peaked Spectrum sources, advection-dominated
sources, starburst galaxies, high-redshift proto-spheroidal
galaxies) to the arcminute scale fluctuations measured by the CBI,
BIMA, and ACBAR experiments are discussed. At 30 GHz, fluctuations
due to radio sources undetected by ancillary low-frequency surveys
may be higher than estimated by the CBI and BIMA groups.
High-redshift dusty galaxies, whose fluctuations may be strongly
enhanced by the effect of clustering, could contribute to the BIMA
excess signal, and dominate at 150 GHz (the ACBAR frequency).
Moreover, in the present data situation, the dust emission of
these high-redshift sources set an unavoidable limit to the
detection of primordial CMB anisotropies at high multipoles, even
at frequencies as low as $\simeq 30$ GHz. It is concluded that the
possibility that the excess power at high multipoles is dominated
by unsubtracted extragalactic sources cannot be ruled out. On the
other hand, there is room for a contribution from the
Sunyaev-Zeldovich effect within clusters of galaxies, with a
density fluctuation amplitude parameter $\sigma_8$ consistent with
the values preferred by current data.}

\maketitle



\section{Introduction}
In the past few years different experiments (BIMA: Dawson et al.
2002; CBI: Mason et al. 2003, Readhead et al. 2004, and ACBAR: Kuo
et al. 2004), aimed at measuring the anisotropies of the cosmic
microwave background (CMB) on arcmin angular scales, have detected
signals at multipoles $\ell > 2000$ in excess of the expected
primordial CMB anisotropies. The origin of this excess signal, in
the range $16\lsim \Delta T\lsim 26\,\mu$K, is not well understood
yet, although several possibilities have been discussed in the
literature.

All experimental groups argue that it cannot be due to
point-source contamination. If so, the most likely candidate is
the thermal Sunyaev-Zeldovich (SZ) effect, which is expected to
dominate CMB anisotropies on angular scales of a few arcminutes
(Gnedin \& Jaffe 2001). However, an interpretation on terms of SZ
effects from clusters of galaxies (Bond et al. 2005; Komatsu \&
Seljak 2002) or to the inhomogeneous plasma distribution during
the formation of large scale structure (Zhang et al. 2002)
requires values of $\sigma_8$ (the rms density fluctuation on a
scale of $8h^{-1}\,$Mpc) significantly higher than indicated by
current data. SZ effects associated with the formation and the
early evolutionary phases of massive spheroidal galaxies could
account for the BIMA signal, although some parameters need to be
stretched to their boundary values (De Zotti et al. 2004).
Alternative interpretations advocate non-standard inflationary
models (Cooray \& Melchiorri 2002; Griffiths et al. 2003).

In this paper we revisit the contributions of extragalactic point
sources to the power spectrum on arcminute scales at the relevant
frequencies, including the possible role of faint sources, with
flux densities too weak to be filtered out, and the effect of
clustering. The outline of the paper is as follows.
In Section 2 we describe the different source populations
which give the dominant contributions to number counts at cm and mm
wavelengths. In Section 3 we present our estimates of arcminute-scale
CMB anisotropies due to extragalactic sources, while in Section 4 we
summarize our main conclusions.

A flat $\Lambda$CDM cosmology with
$\Omega_{\Lambda}$=0.7 has been used throughout the paper.

\section{Extragalactic sources at cm and mm wavelengths}

The estimated contributions of the various populations of
extragalactic sources to the counts at 30 GHz (the frequency of
BIMA and CBI experiments) and at 150 GHz (ACBAR experiment),
obtained from the model of De Zotti et al. (2005), which updates
the model by Toffolatti et al. (1998), are shown in Fig. 1. In
addition to the canonical flat- and steep-spectrum radio sources,
the model takes into account star-forming galaxies with their
complex spectra including both radio (synchrotron plus free-free)
and dust emission, and the source populations characterized by
spectra peaking at high radio frequencies, such as extreme GHz
Peaked Spectrum (GPS) sources and accretion flows on almost
inactive supermassive black-holes in early type galaxies
(ADAF/ADIOS sources).

Because of their inverted low-frequency spectra, GPS and
ADAF/ADIOS sources are potentially worrisome. However, GPS sources
are rare and the analysis made by De Zotti et al. (2000) implies
that they likely have very flat counts and therefore are minor
contributors to small scale fluctuations. Furthermore, the
repeated multifrequency measurements by Tinti et al. (2005) have
demonstrated that most GPS candidates identified with quasars in
the sample of Dallacasa et al. (2000) are in fact flaring blazars,
so that the surface densities of bona-fide GPS sources is probably
substantially lower than estimated by De Zotti et al. (2000), a
conclusion further supported by an examination, carried out by De
Zotti et al. (2005), of GPS candidates in the WMAP sample (Bennett
et al. 2003).

ADAF/ADIOS sources are far more numerous, but have a low radio
power. The estimate by De Zotti et al. (2005) of their counts is
well below that by Perna \& Di Matteo (2000), whose results are
probably affected by a numerical error, and implies that also
these sources do not contribute significantly to the fluctuations
measured by the BIMA and CBI experiments. On the other hand,
Pierpaoli \& Perna (2004; model A) pointed out that if the
standard ADAF model (Narayan \& Yi 1994) is used, these sources
could make up to 40--50\% of the BIMA and CBI excesses. We note,
however, that the standard ADAF scenario faces a number of serious
difficulties, some of which are summarized in Sect. 4.3 of De
Zotti et al. (2005), suggesting that the radio emission is
suppressed by massive outflows. It is therefore likely that the
results of model A by Pierpaoli \& Perna (2004) should be regarded
as, probably generous, upper limits.

As for starburst galaxies, the slope of their differential counts
can exceed 3, if these objects have to account for the very steep
ISOCAM $15\mu$m counts below a few mJy, as implied by recent
analyses (Gruppioni et al. 2003; Franceschini et al. 2003; Pozzi
et al. 2004; Silva et al. 2005). In this case, their main
contribution to small scale fluctuations comes from weak sources,
at $\mu$Jy levels, far fainter than those removed from CBI and
BIMA maps. On the other hand, the counts of such sources are
tightly constrained by $\mu$Jy counts at 1.4 GHz (Richards 2000),
5 GHz (Fomalont et al. 1991), and 8.4 GHz (Fomalont et al. 2002).
Taking such constraints into account and applying an average
spectral index $\alpha=0.8$, appropriate for this class of
sources, we find that they can only provide a minor contribution
to the excess power detected by BIMA and CBI: $\sim
4.3\,\mu\hbox{K}$ at $\ell\simeq 6880$, by using the nominal
$6\sigma$ detection limit, $S_d=150\,\mu$Jy, for BIMA and $\sim
5.0\,\mu\hbox{K}$ at $\ell\simeq 2500$ ($S_d=3.4$ mJy) for CBI,
respectively. Their contribution is negligible at the ACBAR
frequency.

An additional contribution is expected from dusty proto-spheroidal
galaxies, which may account for galaxies selected by SCUBA and
MAMBO surveys (Granato et al. 2001, 2004), whose counts at
$850\,\mu$m and 1.2mm appear to fall down very rapidly at flux
densities above several mJy (Scott et al. 2002; Borys et al. 2003;
Greve et al. 2004). The spectral energy distribution of nearby
dusty galaxies is dominated by synchrotron plus free-free emission
at $\lambda > 2$--3 mm (Bressan et al. 2002), while at shorter
wavelengths dust emission, rapidly raising with frequency ($S_\nu
\propto \nu^4$), takes over. Since these sources are at typical
redshifts $> 2$ (Chapman et al. 2003), dust emission can
significantly contribute to the counts even at 30 GHz. When dust
emission comes in, the counts, already steep because of the effect
of the strong cosmological evolution, are boosted by the large
negative K-correction.

\begin{figure}
  \resizebox{\hsize}{!}{\includegraphics{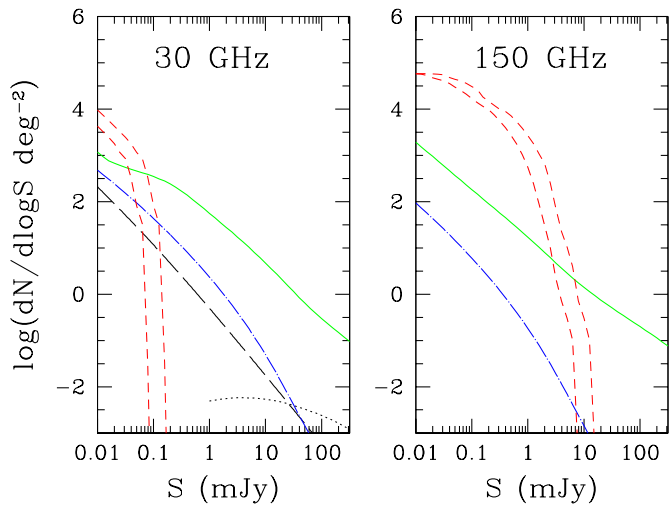}}
  \caption{Differential counts, $dN/d\log S$, of
  different source populations at 30 and 150 GHz, based on the De Zotti et al. (2005)
  evolution model. Solid lines: canonical flat-- plus steep-spectrum
  sources; dot-dashed lines: starburst galaxies; short-dashed
  lines: dusty proto-spheroidal galaxies with (upper curve) and
  without the mm excess (see text). In the left-hand panel only
  the estimated counts of GPS sources (dotted line) and of
  ADAF/ADIOS sources (long-dashed line) are also shown.}
\end{figure}

The poor knowledge of the millimeter emission of these sources,
however, makes estimates of their contributions to the 30 GHz
counts quite uncertain. The two short-dashed lines in Fig. 1 show
the counts we obtain using the physical evolutionary model by
Granato et al. (2004) but with two choices for the spectral energy
distribution (SED). The lower (thicker) line refers to the SED
produced by the code GRASIL (originally described by Silva et al.
1998). An excess emission by a factor $\simeq 2$ at $\lambda \ge
1\,$mm was however detected in several Galactic clouds, combining
Archeops with WMAP and DIRBE data (Bernard et al. 2003; Dupac et
al. 2003), and in NGC1569 (Galliano et al. 2003). The origin of
the excess is still not understood. Possibilities discussed in the
literature are that the grain sizes or composition change in dense
environments or that there is an intrinsic dependence of the dust
emissivity index on temperature (Dupac et al. 2004). If the excess
is due to very cold grains (Reach et al. 1995; Galliano et al.
2003) it cannot be present in the high-$z$ proto-spheroids. But if
it is a general property of the SED of dusty galaxies, the
predicted counts of dusty proto-spheroids are given by the upper
(thin) short-dashed curve.

\begin{figure}[t]
\resizebox{\hsize}{!}{\includegraphics{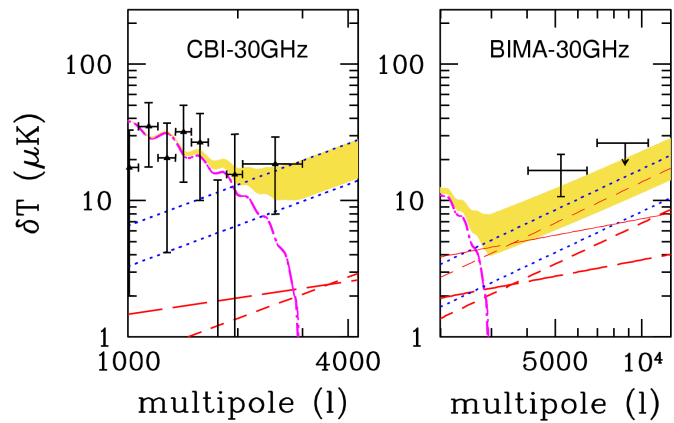}}\label{fig:CBI_BIMA}
\caption{Angular power spectrum ($\delta T =\sqrt{\ell (\ell
+1)C_{\ell}/2\pi}$) measured at 30 GHz by CBI (left-hand panel;
data points from Readhead et al. 2004) and by BIMA (right-hand
panel; data points from Dawson et al. 2002). In each panel, we
have plotted the primordial CMB angular power spectrum
(dot-dashed-line), the estimated range of contributions of
unsubtracted canonical radio sources (dotted lines; see text), and
of Poisson distributed (short-dashed line) and clustered
(long-dashed line) proto-spheroidal galaxies. In the BIMA case,
the upper pair of long- and short-dashed lines refers to
proto-spheroidal galaxies with the submillimeter excess mentioned
in Sect.~2 (not shown in the other panel). The contributions of
the latter sources are insensitive to the adopted flux limit
because of the very steep counts. The shaded areas show the ranges
spanned by the quadratic sum of the different contributions. }

\end{figure}

\section{Contributions of extragalactic sources to arcminute scale anisotropies}

\subsection{Observations with the Cosmic Background Imager (CBI)}

The strategy of the CBI group (Mason et al. 2003; Readhead et al.
2004) to remove the point source contamination comprises pointed
31 GHz observations with the OVRO 40m telescope of all NVSS
sources with 1.4 GHz flux density $\ge 6\,$mJy, and direct counts
at 31 GHz using the CBI deep and mosaic maps. Although the
$4\sigma$ threshold of OVRO observations is 6 mJy, the survey is
99\% complete only at $S_{31\rm GHz} > 21\,$mJy. The limiting flux
density ranges from 6 to 12 mJy in the deep CBI maps, and from 18
to 25 mJy in the mosaic maps. Subtraction of OVRO detected sources
removes two-thirds of the observed power level.

Furthermore, they have adopted the constraint matrix approach to
remove from their dataset all NVSS sources with flux densities
greater than 3.4 mJy at 1.4 GHz (Readhead et al. 2004), and have
estimated the contribution to fluctuations due to sources below
the NVSS cutoff using the observed OVRO-NVSS distribution of
spectral indices and adopting a rather shallow power-law slope for
the counts ($N(>S) \propto S^{-0.875}$); for comparison, Richards
(2000) finds $N(>S) \propto S^{-1.4\pm 0.1}$ for $40\mu\hbox{Jy} <
S_{1.4} < 1\,$mJy (see also Windhorst et al. 1993).

At the flux-density levels relevant for the CBI experiment, apart
from SZ effects (see Sect. 1), the dominant contribution to
fluctuations due to extragalactic sources is expected to come from
the classical steep- and flat--spectrum radio sources. However, an
accurate determination of the 30 GHz fluctuations due to sources
with $S_{1.4} \le 3.4\,$mJy is very difficult because of the
effect of sources with inverted spectra (in fact, Readhead et al.
2004 report the detection of a source, NGC~1068, with $S_{30{\rm
GHz}} \simeq 400\,$mJy, not removed by the constraint matrix), and
of variability. The difficulty is illustrated by the results of
high-frequency surveys. Ricci et al. (2004) found that the 18 GHz
flux densities of extragalactic sources detected by the ATCA pilot
survey are not significantly correlated with the SUMSS flux
densities at 0.84 GHz (i.e. at a frequency not far from that of
the NVSS survey). Waldram et al. (2003) also reported a large
spread (about a factor of 10) of the 15 to 1.4 GHz flux density
ratios of sources detected in their 9C survey at 15 GHz, although
the flux densities at the two frequencies are correlated. They
also noted that pointed 15 GHz observations of the NVSS sources
with $S_{1.4} \ge 25\,$mJy in the area covered by their survey
 would have detected 434 sources above the 9C survey limit of 25
mJy but would have missed 31 sources having $S_{15} \ge 25\,$mJy
but $S_{1.4} < 25\,$mJy. The distribution of spectral indices has
a systematic drift towards flatter values with decreasing
low-frequency flux density down to $S_{1.4} \simeq 1\,$mJy
(Windhorst et al. 1993), so that the fraction of sources with
inverted spectra is expected to be higher at the fainter flux
density levels of interest here.

High frequency surveys emphasize flat-spectrum sources. The
dominant flat-spectrum population are blazars, that are highly
variable on timescales of years and whose variability amplitude
increases with frequency (Impey \& Neugebauer 1988; Ciaramella et
al. 2004). The monitoring campaigns at 22, 37 and 87 GHz by the
Mets\"ahovi group (Ter\"asranta et al. 1998) have shown that
intensity variations by factors of several are common at these
frequencies, so that a substantial fraction of such sources may
have had, at the moment of the CBI observations, 30 GHz fluxes
higher than $3.4\,$mJy, even by a considerable factor. Variability
can indeed account, to a large extent, for the lack of a
correlation between the ATCA 18 GHz and the SUMSS 0.84 GHz flux
densities (Ricci et al. 2004) and for the large spread of the 15
to 1.4 GHz flux density ratios (Waldram et al. 2003).

To appraise residual fluctuations at 30 GHz due to unsubtracted
sources we have adopted the analytical description of the counts
below a few mJy by Richards (2000; $dN/dS_{1.4} = A
S_{1.4}^{-\gamma}$ with $A= 8.3 \pm 0.4$ and $\gamma= 2.4 \pm
0.1$) and computed the Poisson fluctuations at 30 GHz of those
sources with $S_{1.4} < 3.4\,$mJy, assuming a Gaussian
distribution of spectral indices with mean $\bar{\alpha} = 0.4$
(Fomalont et al. 1991; Windhorst et al. 1993) and two values of
the dispersion ($\sigma = 0.3$ or 0.4), based on the width of the
distribution of $\alpha_{1.4}^{15.2}$ of Waldram et al. (2003,
their Fig. 9). A comparison with the 30 GHz counts yielded by the
De Zotti et al. (2005) model, which takes into account the
available information from high-frequency surveys, shows that, in
the 30 GHz flux density range relevant to estimate fluctuations,
the counts extrapolated using the upper values of $A$ and $\gamma$
are somewhat too high if $\sigma = 0.4$; more consistent counts
(only slightly above the model predictions) are obtained with the
central values $A=8.3$, $\gamma= 2.4$. To bound the plausible
range of residual fluctuations we have therefore considered the
cases $A=8.3$, $\gamma= 2.4$, $\sigma = 0.4$ (upper dotted line in
the left-hand panel of Fig.~\ref{fig:CBI_BIMA}) and $A=7.9$,
$\gamma= 2.3$, $\sigma = 0.3$ (lower dotted line).

The additional contribution to CMB fluctuations given by
correlated positions in the sky of canonical steep- and
flat-spectrum radio sources has been recently analyzed by
Gonz\'alez-Nuevo et al. (2005). Their outcomes indicate that the
extra power due to the clustering of radio sources cannot, by
itself, explain the excess signal detected by CBI and BIMA. Using
the $w(\theta)$ estimated by Blake \& Wall (2002) from sources in
the NVSS survey down to $S\simeq 10$ mJy - which can represent a
realistic approximation to the clustering properties of faint
undetected sources in the CBI fields - they found that clustered
radio sources at $S_{30{\rm GHz}}< 3.4$ mJy can give an extra
power $\Delta T\simeq 3$-$4$ $\mu$K, which has to be summed up -
in quadrature - to the Poisson term, $\Delta T\simeq 20$-22
$\mu$K. The dominance of Poisson over clustering fluctuations even
at faint fluxes is due to the strong dilution of the clustering
signal of extragalactic radio sources by the broadness of their
luminosity function and of their redshift distribution (Dunlop \&
Peacock 1990; Toffolatti et al. 1998; Negrello et al. 2004).


\subsection{Observations with the Berkeley-Illinois-Maryland Association Array (BIMA)}


To remove the point source contamination, the BIMA group (Dawson
et al. 2002) have carried out a VLA survey at 4.8 GHz of their
fields. These observations reached a rms flux of $\simeq
25\,\mu$Jy$\,\hbox{beam}^{-1}$ for a $9'$ FWHM region with center
coinciding with the center of the corresponding BIMA field.
Sources with flux density $> 6\sigma_{VLA}$ within $8'$ of the
pointing center were projected out. On the other hand, point
sources with flux densities $S_{4.8}> 150\ \mu$Jy, lying at an
angular distance $\theta$ from the BIMA field center, cannot be
detected (and removed) by VLA observations if
$S_{4.8}<S_{lim}(\theta)=6\sigma/f(\theta)$, where $f(\theta)$ is
the VLA response function, assumed Gaussian.

Therefore, we have estimated the fluctuations in the BIMA fields
due to sources fainter than $150\,\mu$Jy$/f(\theta)$, where
$\theta$ is the angular distance from the pointing direction, by
adopting the number counts at 4.8 GHz of Fomalont et al. (1991),
$N(>S)=(23.2\pm 2.8) S_{4.8,\mu{\rm Jy}}^{-1.18\pm 0.19}$
arcmin$^{-2}$, extrapolated to 30 GHz with a mean spectral index
$\alpha=0.4$ ($S_\nu \propto \nu^{-\alpha}$), appropriate for the
relevant flux-density range (Fomalont et al. 1991; Windhorst et
al. 1993). The corresponding power spectrum is shown in Fig.~2
(right-hand panel), where the shaded area reflects the range of
values corresponding to the uncertainties in the counts of
Fomalont et al. (1991) and in the dust emission spectrum of high
redshift spheroids (see below).

The fluctuations due to forming spheroidal galaxies, not
represented in the 4.8 GHz counts, get comparable contributions
from both the Poisson and the clustering term, while the latter
term turns out to be small, compared to the former, for the other
classes of sources relevant here. Adopting the standard expression
for the two-point correlation function, $\xi(r)=(r/r_0)^{-1.8}$
with a constant comoving clustering length
$r_0=8.3\hbox{h}^{-1}\,$Mpc, h being the Hubble constant in units
of $100\,\hbox{km}\,\hbox{s}^{-1}\,\hbox{Mpc}^{-1}$ (see Negrello
et al. 2004), we find, at $\ell_{\rm eff} = 6864$, a Poisson
contribution of $\simeq 5\mu$K and a clustering contribution of
$\simeq 3\mu$K. Clearly these contributions, to be summed in
quadrature to the contribution discussed above, have a minor
effect. If, however, these sources show the mm excess mentioned in
Sect. 2, their contribution to fluctuations would be approximately
doubled (see Fig. 2), and, summed in quadrature to the above
estimate of the contribution of radio sources,  could account for
the reported excess signal.

\subsection{Observations with the Arcminute Cosmology Bolometer Array Receiver (ACBAR)}

The ACBAR measurements in the 150 GHz band reported by Kuo et al.
(2004) up to multipoles $\ell = 3000$ are consistent with the
primordial CMB power spectrum predicted by standard cosmological
models. On the other hand, the measured power in the highest
multipole bin is also consistent with the excess detected by the
CBI experiment. Fluctuations due to extragalactic radio sources in
the ACBAR band are quite small (of a few $\mu$K at $\ell_{\rm
eff}=2507$). Quoting results by Blain et al. (1998), Kuo et al.
(2004) conclude that dusty proto-galaxies are also not expected to
contribute significantly to the observed signal. However, the
Blain et al. (1998) estimate actually refers to Poisson
fluctuations only, while it has been pointed out by several
authors (Scott \& White 1999; Haiman \& Knox 2000; Magliocchetti
et al. 2001; Perrotta et al. 2003; Negrello et al. 2004) that the
strong positional correlation of SCUBA galaxies, indicated by
observational data and theoretical arguments, implies that the
main contribution on ACBAR scales comes from source clustering. On
the other hand, the rest-frame mm-wave emission of these sources
is poorly known.  If dusty proto-spheroidal galaxies have the
excess emission (compared to model expectations) discussed in
Sect. 3.2 (Fig.~3, right-hand panel), the power measured in the
highest multipole bin is easily explained in terms of fluctuations
due to undetected high-$z$ galaxies. Anyway, dusty proto-spheroids
are always giving a relevant contribution to the measured power,
even without excess emission (Fig.~3, left-hand panel).

\begin{figure}
  \resizebox{\hsize}{!}{\includegraphics{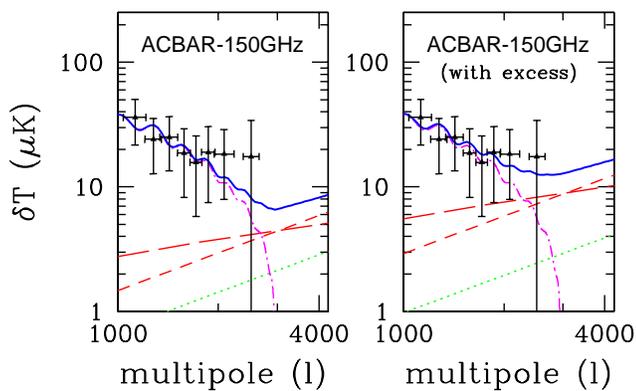}}
  \caption{Angular power spectrum ($\delta T =\sqrt{\ell (\ell +1)
  C_{\ell}/2\pi}$) measured by ACBAR at 150 GHz (data
  points are from Kuo et al. 2004). The dotted, long- and short-dashed,
  and dot-dashed curves have the same meaning as in Fig.~2; the solid
  lines are the quadratic sum of all the contributions. The contribution given by
  undetected Poisson distributed and clustered proto-spheroidal
  galaxies has been calculated without (left-hand panel) and with
  (right-hand panel) the excess mm-wave emission discussed in the text.
  In this case we adopted $S_{\rm lim}\simeq 24$ mJy (left-hand panel)
  and $S_{\rm lim}\simeq 43$ mJy (right-hand panel) for source detection.
  These limits correspond to the $5\sigma$ source detection threshold
  estimated as in Negrello et al. (2004) for a 5$^\prime$ FWHM.
  }
\end{figure}

\section{Conclusions}

Contamination from extragalactic point sources appears to be a
likely candidate to account for a large fraction, perhaps for most
of the excess power on arcminute scales detected by the CBI and
BIMA experiments. The fluctuations due to radio sources undetected
by the ancillary low-frequency surveys may in fact be higher than
estimated by the CBI and BIMA groups. On the other hand, we argue
that extreme GHz Peaked Spectrum sources and advection dominated
sources, potentially worrisome because of their spectra peaking at
high microwave/mm-wave wavelengths, should provide only a minor
contribution to the CBI and BIMA signals.

Although the very steep $15\,\mu$m counts of starburst galaxies
below a few mJy imply a very strong cosmological evolution, the
radio surveys down to $\mu$Jy levels constrain the contribution of
their {\it radio} emission to fluctuations at 30 GHz to be
relatively small. On the other hand, the {\it dust} emission at
rest-frame mm wavelengths from star-forming galaxies at
high-redshifts, such as those detected by SCUBA and MAMBO surveys,
can be redshifted down to 30 GHz. Using the spectral energy
distributions given by the Granato et al. (2004) model, we find
that these sources yield fluctuations of a few to several $\mu$K
on arcminute scales. Their rest-frame spectral energy distribution
at mm wavelengths, however, is poorly known, and may well be
higher than implied by the adopted model.

Moreover, observational indications and the theoretical arguments
converge in suggesting that they are highly clustered (see
Negrello et al. 2004, and references therein), so that their
fluctuations may be strongly super-Poissonian for multipoles $\ell
\lsim 3000$ (remember that the clustering-to-Poisson ratio
increases with the angular scale, i.e. with decreasing multipole
number, De Zotti et al. 1996). Clustering fluctuations of the
high-$z$ galaxies detected by (sub)-mm SCUBA and MAMBO surveys may
indeed dominate the contamination by extragalactic sources of the
signal measured by the ACBAR experiment at 150 GHz.

Because the dust emission spectrum rises very steeply with
frequency, lower frequency surveys cannot be used to remove their
effect from 30 GHz maps. In the present data situation, they
therefore set an {\it unavoidable} limit to the determination of
the primordial CMB angular power spectrum at high multipoles, even
at frequencies as low as $\sim 30$ GHz.

We stress that the present results are fully compatible with the
estimated contributions of Sunyaev-Zeldovich effects in clusters
of galaxies to the arcminute scale anisotropies. In fact, while an
interpretation of the full CBI and BIMA signals in terms of SZ
fluctuations would require a density fluctuation amplitude
(measured by the parameter $\sigma_8$) at or above the limit
allowed by current data (Bond et al. 2005), our analysis leaves
room for an SZ contribution corresponding to the $\sigma_8$ values
favoured by analyses of CMB, cosmic shear, and large scale
structure data (Spergel et al. 2003; Pierpaoli et al. 2003; Van
Waerbecke et al. 2005).

New interesting constraints on the CMB angular power spectrum up
to $\ell\sim 2500$ at 34 GHz should be provided in the near future
by the VSA experiment. The reduced noise level of the new
configuration and an effective cleaning of deep fields down to
$\sim 5$ mJy -- by dedicated observations with the Ryle Telescope
at 15 GHz -- will shed new light on the nature of the excess at
high multipole and on the point source populations mainly
contributing to the number counts at $S_{34}\sim$ a few mJy.
Moreover, {\sl Planck} HFI data as well as the forthcoming surveys
by the {\sl Herschel} telescope -- at frequencies where the
emission due to cold dust grains is the dominant one -- shall be
unique in determining much better the cosmological evolution, the
emission and the clustering properties of high-redshift dusty
galaxies.

\begin{acknowledgements}

We are grateful to K. Dawson and to G. Holder for very stimulating
comments and clarifications on the point source subtraction for
the BIMA experiment, and to the referee whose comments greatly
helped improving the paper. LT, JGN and FA thank the Spanish MEC
(Ministerio de Educaci\'on y Ciencia) for partial financial
support (project ESP2004-07067-C03-01). JGN acknowledges a FPU
fellowship and an ``Ayuda'' for Short Research Periods of the
Spanish Ministry of Education (MEC). JGN also thanks the
SISSA-ISAS, International School for Advanced Studies (Trieste,
Italy), where his share of this work was completed, for the warm
hospitality.

\end{acknowledgements}

\end{document}